# Cold Isostatic Pressing to improve the mechanical performance of additively manufactured metallic components

I.I. Cuesta [(1)], E. Martínez-Pañeda [(2)], A. Díaz [(1)], J.M. Alegre [(1)]

[(1)] Structural Integrity Group, Universidad de Burgos, Avda. Cantabria s/n, 09006 Burgos. SPAIN
[(2)] University of Cambridge, Department of Engineering, Trumpington Street, Cambridge CB2 1PZ, UNITED KINGDOM

Telephone: +34 947 258922; e-mail: iicuesta@ubu.es

**Abstract**

Additive Manufacturing is becoming a technique with great prospects for the production of components with new designs or shapes that are difficult to obtain by conventional manufacturing methods. One of the most promising techniques for printing metallic components is Binder Jetting, due to its time efficiency and its ability to generate complex parts. In this process, a liquid binding agent is selectively deposited to adhere the powder particles of the printing material. Once the metallic piece is generated, it undergoes a subsequent process of curing and sintering to increase its density (Hot Isostatic Pressing). In this work, we propose subjecting the manufactured component to an additional post-processing treatment involving the application of a high hydrostatic pressure (5000 bar) at room temperature. This post-processing technique, so-called Cold Isostatic Pressing (CIP), is shown to increase the yield load and the maximum carrying capacity of an additively manufactured AISI 316L stainless steel. The mechanical properties, with and without CIP processing, are estimated by means of the Small Punch Test, a suitable experimental technique to assess the mechanical response of small samples. In addition, we investigate the porosity and microstructure of the material according to the orientations of layer deposition during the manufacturing process. Our observations reveal a homogeneous distribution independent of these orientations, evidencing thus an isotropic behaviour of the material.

# 1. Introduction

Additive Manufacturing (AM) is experiencing an increasing popularity in both academic and industrial applications; see the work by Frazier [1] for a review. Its versatility in manufacturing engineering components by metal deposition is making AM a feasible alternative for the production of parts and prototypes in different sectors, such as the biomedicine, aerospace or automotive [2-4]. One advantage of AM is the reduction in the time elapsed from the conception of the component to its final form, as it does not require the design and manufacturing of special tools unlike other production processes like casting or forming. Among the different AM techniques proposed, Binder Jetting is gaining particular traction due to its time efficiency and its ability to generate complex components. In this process, an inkjet print head selectively deposits a liquid binding material across a bed of powder. Thus, the material layers are superimposed to form the desired part while the print nozzle strategically drops the binding agent into the powder surface. Once the metallic piece is generated, it undergoes a subsequent process of curing and sintering (Hot Isostatic Pressing - HIP) to achieve the desired density. The benefits of a HIP post-processing step in improving the mechanical properties of additively manufactured materials have been documented at large. For example, Dadbakhsh and Hao [5] examined the role of HIP in Al composite parts generated by selective laser melting (SLM), finding an increase in density and a decrease in hardness. Similarly, Srivastava *et al.* [6] achieved a remarkable increase in density by applying HIP to a bulk metallic glass cast component. In addition, AlMangour *et al.* [7] showed that HIP is an effective post-treatment technology for suppressing larger pores and fusion defects from additively manufactured components. Important physical insight into the benefits of HIP post-processing has also been gained from the numerical perspective. For example, Kim [8] used numerical creep techniques to model the HIP process in an AISI 316L steel by incorporating the numerous diffusion mechanisms taking place into a novel constitutive model. However, the HIP post-process is typically applied in isolation, and the benefits of combining HIP with other post-treatments have been scarcely explored in metal additive manufacturing.

In this paper, we propose and investigate the effect of applying a high hydrostatic pressure as an additional post-processing technique to further increase material density and mechanical performance. The so-called Cold Isostatic Pressing (CIP) technique is used to subject metallic samples to high pressures at room temperature. CIP post-proccesing has been used in combination with HIP in the context of conventional powder metallurgy, see for example the pioneering work by Ng *et al.* [9], but its potential in additive manufacturing remains to be explored. We aim at filling this knowledge gap by applying very high pressures, up to 6000 bar, by means of a new device, recently patented, which has been developed based on High Pressure Processing (HPP) technology. More specifically, we evaluate the mechanical properties of additively manufactured AISI 316L steel samples that have been produced by means of Binder Jetting and subsequently subjected to HIP and CIP at 5000 bar. It is expected that CIP post-processing will compact the samples, reducing the number and shape of internal defects. The capabilities of CIP in improving material performance are assessed by means of the Small Punch Test (SPT). The SPT was initially developed by Baik *et al.* [10] to study the influence of radiation on the ductile-to-brittle transition temperature in metallic materials. Since then, it has been successfully employed to measure both mechanical and fracture properties from small samples; see the works by Saucedo-Muñoz *et al.* [11] and Ju *et al.* [12] for fracture toughness estimations, and the articles by Alegre *et al.* [13] and Hou *et al.* [14] for creep properties. The SPT is particularly suitable for this application, as the pressurised volume inside the cylindrical CIP device is not sufficiently large to accommodate conventional test specimens. In addition to investigating the effect of CIP post-processing in AM steels, we analyse porosity and material microstructure to assess the influence of layer orientation during manufacturing process.

The remainder of this manuscript is structured as follows. Section 2 describes the material employed and the microstructure analysis conducted. In Section 3, we provide details of the CIP process and the subsequent small-scale mechanical tests conducted by means of the SPT. The results are presented and discussed in Section 4. Finally, concluding remarks are given in Section 5.

## 2. Material

The present study is conducted on a stainless steel studied by Nastac *et al.* [15], AISI 316L, which has been additively manufactured by means of the Binder Jetting method in an ExOne M-Flex metal 3D printer. The chemical composition and mechanical properties of the 316L alloy are given in Tables 1 and 2, respectively.

The microstructure analysis reveals the existence of numerous pores. A representative Scanning Electron Microscope (SEM) micrograph is shown in Figure 1, in which the size of the pores and the porosity distribution can be clearly observed; arrows are used to help the eye. In addition, we investigate possible anisotropies by conducting a microstructure analysis along the three characteristic orientations intrinsic to the Binder Jetting process: L, T, and S. These are defined as follows:

- L. Orientation defined by the advance direction of the printhead.
- T. Orientation defined by the direction perpendicular to the advance of the printhead, i.e. normal to the L orientation.
- S. Orientation perpendicular to the plane LT and coincident with the vertical movement of the printing bed.

Figure 2 shows the microstructure found for each of the planes defined by the orientations L, T and S, i.e. LT, LS and TS. Different sizes and shapes of the pores can be observed as well as the grain pattern; the grain size is found to range between 30 and 70 µm. It can be readily seen that the number of pores is similar in every plane and that the distribution is relatively uniform. Hence, an isotropic behaviour can be assumed and one can neglect the layering effects during Binder Jetting.

## 3. Methodology

### 3.1 Cold Isostatic Pressing

As described schematically in Figure 3, an HPP-based device is employed to carry out the high hydrostatic pressure post-processing at room temperature. This device is based on a conveniently modified high pressure intensifier that is coupled to a universal testing machine. Thus, the device is connected to a universal testing machine MTS 810 with a load frame capacity of 250 kN by two jaws, at the top and the bottom. The device comprises a bearing tube inside which a high-pressure cylinder is housed, through which a stem passes. A flange crossed by the stem is arranged at the lower end of the high-pressure cylinder. At the upper end there is a stopper that closes the assembly, leaving an interior space where the miniature specimens that are going to be pressurized are housed. The lower jaw transmits the vertical movement to the stem and pressurises the inner chamber of the cylinder.

The Cold Isostatic Pressing (CIP) procedure is performed in three steps: first, the hydrostatic pressure is raised up to 5000 bar, then it is maintained for 3 minutes and, finally, the pressure is removed in a controlled manner. Once the pressure unloading has finished, the device is disassembled and the specimens can be collected and tested.

*3.2 Small Punch Testing*

The mechanical assessment of the effect of the CIP process on the material properties is conducted by means of the Small Punch Test (SPT). Extensive details of the testing equipment and the experimental procedure are given in the SPT CEN code of practice [16]. As described in Figure 4, the SPT consists on punching a small specimen with its outer edges embedded by two dies. Small Punch Tests have been carried out on a universal testing machine MTS Criterion 43 with 10 kN load capacity. We employ lubrication to minimize the influence of friction, see details in the works by Cuesta *et al.* [17] and Martínez-Pañeda *et al.* [18]. The punch displacement and the corresponding applied load are recorded during the test, being the resulting load-displacement curve the main outcome of the SPT experiment. As proposed by Martínez-Pañeda *et al.* [19], the load-displacement curve can be divided into different zones, each influenced by the characteristic elastic-plastic parameters of the material. Thus, from the punch load versus displacement curve, one can obtain a number of parameters that can then be correlated with the nominal properties of the material. Of particular interest for this investigation are the load at the onset of yield $P_y$, the maximum value of the load $P_{max}$, and the displacement at maximum load $\Delta_{P_{max}}$.

One of the most influencing works on SPT correlations for determining the mechanical properties is the one by Mao and Takahashi [20]. They established a relationship between the yield load from the SPT, $P_y$, and the material yield stress, $\sigma_y$, through the empirical equation:

$$\sigma_y = \alpha \cdot \frac{P_y}{t^2} \tag{1}$$

Here, $t$ is the thickness of the SPT sample (usually, 0.5 mm), and $\alpha$ is a non-dimensional empirical coefficient that is characteristic of each material. In steels, $\alpha = 360$. Different fitting strategies have been proposed by García [21] for the determination of the yield load $P_y$, i.e. the load delimiting zone I and zone II in the SPT typical curve; we choose to use here the so-called offset method.

SPT specimens have been extracted from a component printed by the Binder Jetting method, obtaining square specimens of $10 \times 10 \, mm^2$ in each of the defined planes: LT, LS and TS. The small specimens were polished to achieve a uniform thickness of roughly 0.5 mm. SPT experiments are carried out, at room temperature, for 316L samples with and without CIP post-processing. The tests are quasi-static, with the punch displacement rate being equal to $v = 0.5$ mm/min. The punch diameter equals $d_p = 2.5$ mm, whereas the size of the lower die is characterized by a diameter of $D_d = 4$ mm and a round radius $r = 0.5$ mm. It is important to emphasize that, due to the specimen dimensions, the CIP post-processing does not solely lead to volumetric strains.

The resulting force versus displacement curves obtained from the experiment are normalized by the measured specimen thickness $t$ to account for small differences in thicknesses that may have arisen after polishing. With this objective, following Cuesta *et al.* [22], we determine an effective load $P_{0.5}$ from the experimentally measured load $P_{test}$ and the actual specimen thickness $t$. The normalisation procedure is divided on two stages that meet at the inflection point of the curve, where zones II and III of the load-displacement curve intersect. Thus, when $P_{test}$ is smaller than the load at the inflection point, $P_{INF}$, the effective load is given by

$$P_{0.5} = 0.5^2 \cdot \frac{P_{test}}{t^2} \qquad P_{test} < P_{INF} \tag{2}$$

While the following expression is employed for load normalisation after this point:

$$P_{0.5} = 0.5 \cdot \frac{P_{test}}{t} + 0.5 \cdot \frac{P_{INF} \cdot (0.5 - t)}{t^2} \qquad P_{test} > P_{INF} \tag{3}$$

To ensure reproducibility, three experiments have been conducted for every combination of specimen orientation (L, T, S) and post-processing treatment (with and without CIP).

## 4. Results and discussion

The values measured with the Small Punch Test (SPT) for the load at yield $P_y$, the maximum value of the load $P_{max}$, and the displacement at maximum load $\Delta_{P_{max}}$, are listed in Tables 3 and 4.

Specifically, Table 3 shows the results obtained for the samples where CIP post-processing has not been conducted while in Table 4 the results are listed for the samples subjected to CIP; in both cases, average values are also provided for each characteristic parameter of the SPT. The results reveal noticeable differences between the two scenarios. Namely, applying a CIP treatment at 5000 bar leads to an increase in the yield load, the maximum load and the displacement at $P_{max}$ of, respectively, 6.5% ($P_y$), 3.1% ($P_{max}$), and 1.4% ($\Delta_{P_{max}}$). The significant increase in yield stress and ultimate strength observed suggests that CIP treatments at very high pressures can translate into an improvement in the mechanical properties of additively manufactured steels. The application of a high hydrostatic pressure introduces plastic deformation in the vicinity of the voids, influencing their shape and size; this has been examined in great detail by Sket *et al.* [23]. Sket and co-workers observed an enhancement of about 30.5 MPa in the uniaxial stress over the plastic region in a Mg AZ91 Alloy with initial yield stress of approximately 90 MPa. This larger sensitivity to the effect high hydrostatic pressures agrees with expectations, due to the different mechanical properties of magnesium alloys and stainless steels

The representative load versus displacement curves obtained from the SPT are shown in Figure 5 for the two scenarios considered – with and without CIP processing. The curves follow each other closely until the vicinity of the maximum load, where the sample without CIP - which is expected to have larger pores - exhibits an earlier failure and a smaller maximum load carrying capacity. The failure mechanism is the same in both cases: ductile damage characterized by the circumferential fracture shown in the SEM image included as inset in Figure 5. We conclude that the use of CIP post-processing to improve the mechanical performance of AM steels does not bring qualitative changes but renders significant quantitative improvements in enhancing yielding and damage resistance.

Remarkably, two of the tested specimens suffered a premature failure in orientation L, which is defined by the advance direction of the printhead. The two samples, B3 and E3, correspond to the LS plane. Figure 6 shows the load-displacement curve of both specimens including also the corresponding SEM images of the observed fracture. The load versus displacement curves exhibit evident differences with the results shown in Figure 5. With the aim of determining the micromechanical origin of this different shape, different scales of the fracture surface of specimen E3 are shown in Figure 7, where a mix between ductile and intergranular failure mechanisms can be observed for the L orientation. This fracture aspect is substantially different from that observed in every other specimen, where a ductile fracture surface, populated with numerous dimples, is typically observed; see Figure 8. The fractured area shown in Figure 7 exhibits a longitudinal aspect that appears to coincide with the L direction of advance of the printhead. Thus, the anomalous behaviour associated with samples with LS orientation is likely to be intrinsically related with the lack of adhesion between binder jetting layers.

## 5. Conclusions

We propose and assess the use of Cold Isostatic Pressing (CIP) post-processing techniques to compact samples of AISI 316L steel that have been additively manufactured (AM) by Binder Jetting. Pressures of 5000 bar are applied by means of a novel device that builds upon High Pressure Processing (HPP) technology; the aim is to improve the mechanical response of the AM specimens. Due to the limited size of the samples that can be accommodated in this device, the Small Punch Test is employed to characterize material performance with and without CIP post-processing. Specifically, three material parameters are extracted: the yield load $P_y$, the maximum load carrying capacity $P_{max}$, and the failure displacement $\Delta_{P_{max}}$. In addition, a microstructure analysis is conducted to identify potential anisotropies that may arise as a consequence of the Binder Jetting manufacturing process.

Our main findings are twofold. First, little differences are observed in the microstructure along the three characteristic planes intrinsic to the Binder Jetting Process. However, mechanical testing reveals early cracking and cleavage-like features in the LS plane samples. Second, the use of CIP post-processing appears to improve the mechanical performance of AM steels. Small Punch Test measurements show that, when the CIP technique is employed as an additional post-processing of the material, there is a 6.5% increase in the yield resistance, a 3.1% increase in the critical load, and a 1.4% increase in the displacement to failure. In addition, this enhancement due to CIP appears to be sensitive to the characteristic orientations intrinsic to the Binder Jetting process. Future work will involve gaining further insight into this effect.

## 6. Acknowledgments

The authors wish to thank the funding received from the Ministry of Education of the Regional Government of Castile and Leon under the auspices of the support for the Recognized Research Groups of public universities of Castile and Leon started in 2018, Project: BU033G18. The SEM

*Table 1. Chemical composition of AISI 316L steel studied by Nastac et al. [15].*

| Element | Cr | Ni | Mo | Mn | Si | P | C | S | Fe |
|---------|------|------|------|------|------|------|------|------|------|
| wt. % | 16.0-18.0 | 10.0-14.0 | 2.0-3.0 | Max 2 | Max 1 | Max 0.04 | Max 0.03 | Max 0.03 | balance |

*Table 2. Mechanical properties of AISI 316L steel studied by Nastac et al. [12].*

| Yield strength $\sigma_Y$ (MPa) | Ultimate tensile strength $\sigma_{UTS}$ (MPa) | Elongation at break, % | Hardness (HRB) | Density (g/$cm^3$) |
|---|---|---|---|---|
| 214 | 517 | 43 | 66 | 7.7 |

Table 3. Characteristic SPT parameters $P_y$, $P_{max}$ and $\Delta_{P_{max}}$ for the SPT specimens without CIP.

| Orientation | Specimen | $P_y$ (kN) | $P_{max}$ (kN) | $\Delta_{P_{max}}$ (mm) |
|---|---|---|---|---|
| TS | A1 | 0.193 | 1.749 | 2.147 |
|  | A2 | 0.195 | 1.817 | 2.073 |
|  | A3 | 0.206 | 1.883 | 2.210 |
|  | A4 | 0.209 | 1.878 | 2.101 |
|  | A5 | 0.205 | 1.801 | 1.985 |
| LS | B1 | 0.214 | 1.804 | 2.213 |
|  | B2 | 0.185 | 1.851 | 2.226 |
|  | B3 | - | - | - |
| LT | C1 | 0.201 | 1.807 | 2.257 |
|  | C2 | 0.204 | 1.914 | 2.286 |
|  | C3 | 0.205 | 1.914 | 2.225 |
| *Average:* |  | **0.202** ±0.008 | **1.841** ±0.055 | **2.170** ±0.094 |

Table 4. Characteristic SPT parameters $P_y$, $P_{max}$ and $\Delta_{P_{max}}$ for the SPT specimens with CIP.

| Orientation | Specimen | $P_y$ (kN) | $P_{max}$ (kN) | $\Delta_{P_{max}}$ (mm) |
|---|---|---|---|---|
| TS | D1 | 0.212 | 1.945 | 2.268 |
|  | D2 | 0.215 | 1.828 | 2.214 |
|  | D3 | 0.231 | 1.840 | 2.159 |
|  | D4 | 0.233 | 1.929 | 2.163 |
| LS | E1 | 0.208 | 1.842 | 2.156 |
|  | E2 | 0.203 | 1.952 | 2.259 |
|  | E3 | - | - | - |
| LT | F1 | 0.209 | 1.808 | 2.180 |
|  | F2 | 0.212 | 1.943 | 2.180 |
|  | F3 | 0.217 | 2.061 | 2.316 |
| *Average:* | | **0.215** ±**0.010** | **1.904** ±**0.082** | **2.210** ±**0.058** |

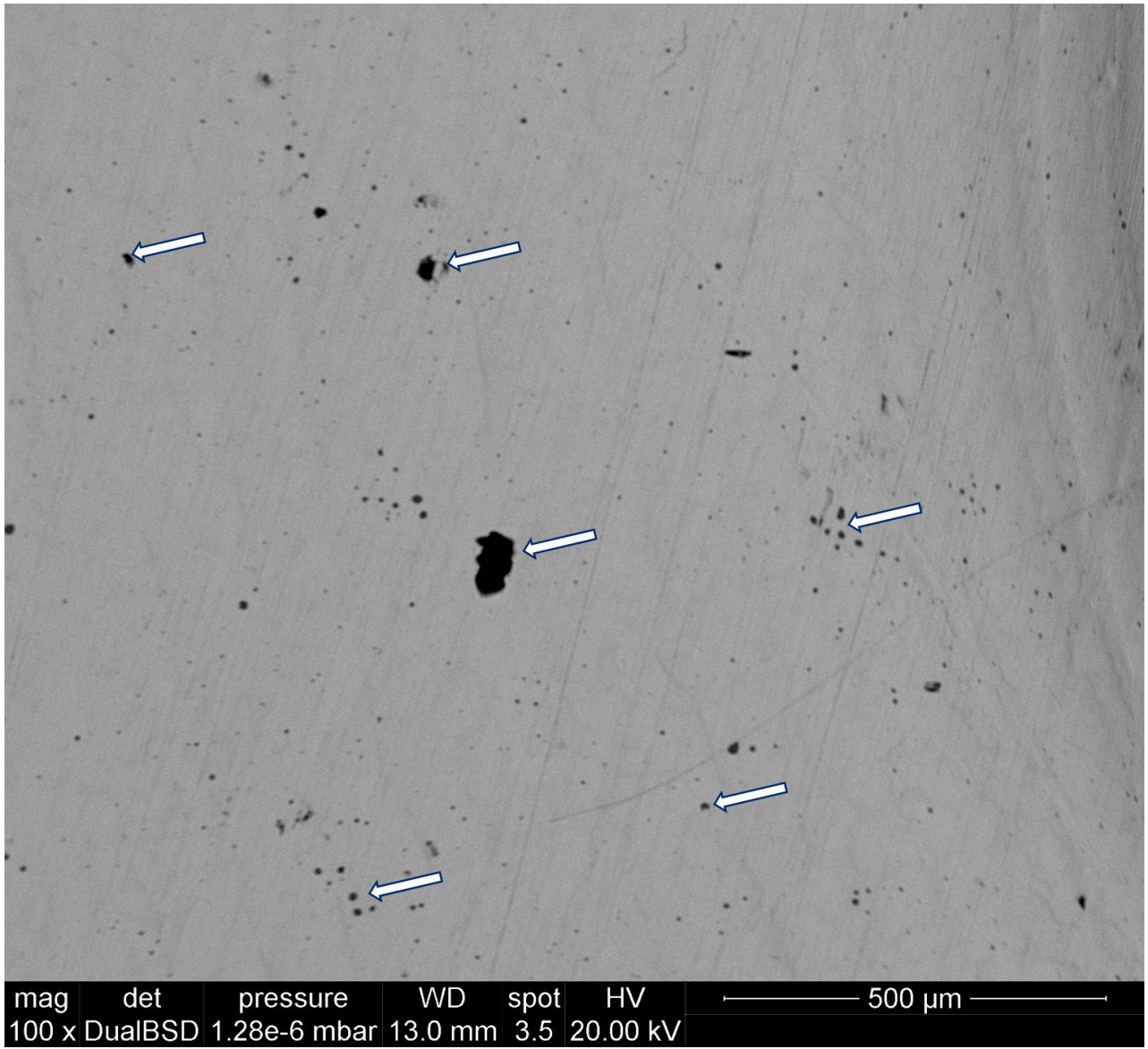

*Figure 1. Porosity distribution in AISI 316L manufactured using Binder Jetting.*

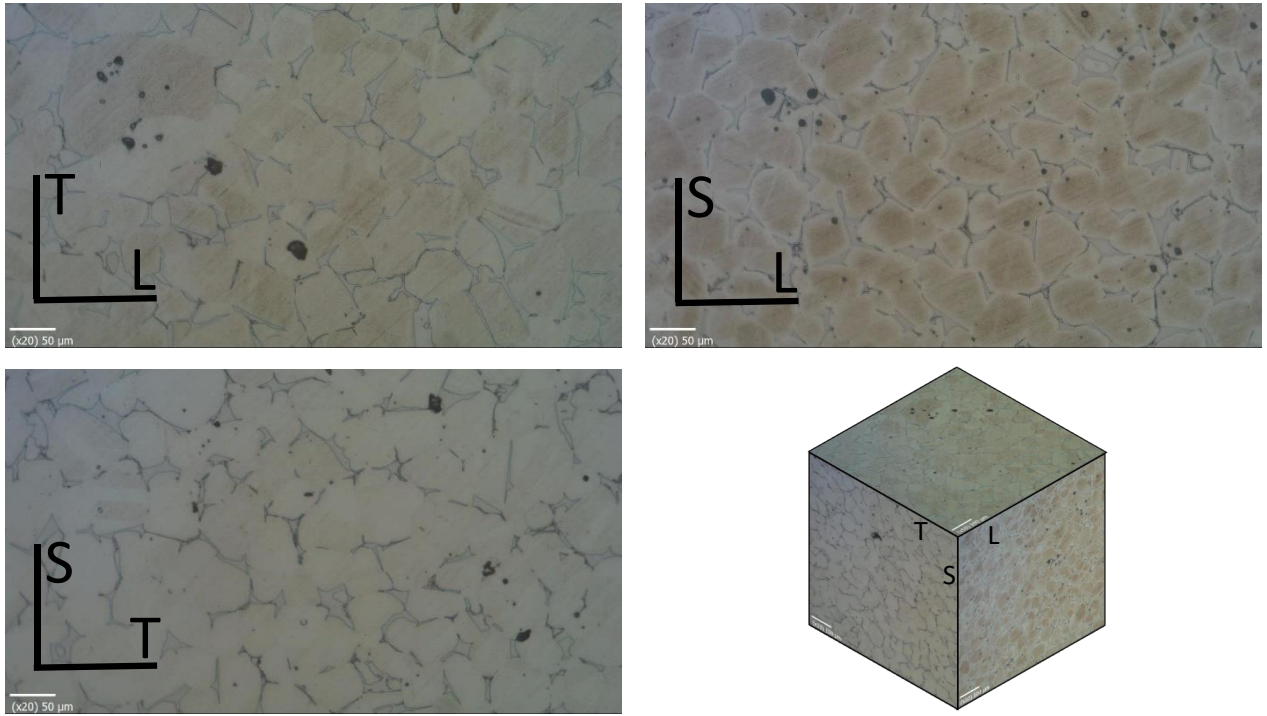

*Figure 2. Orientations and microstructure in AISI 316L manufactured using Binder Jetting.*

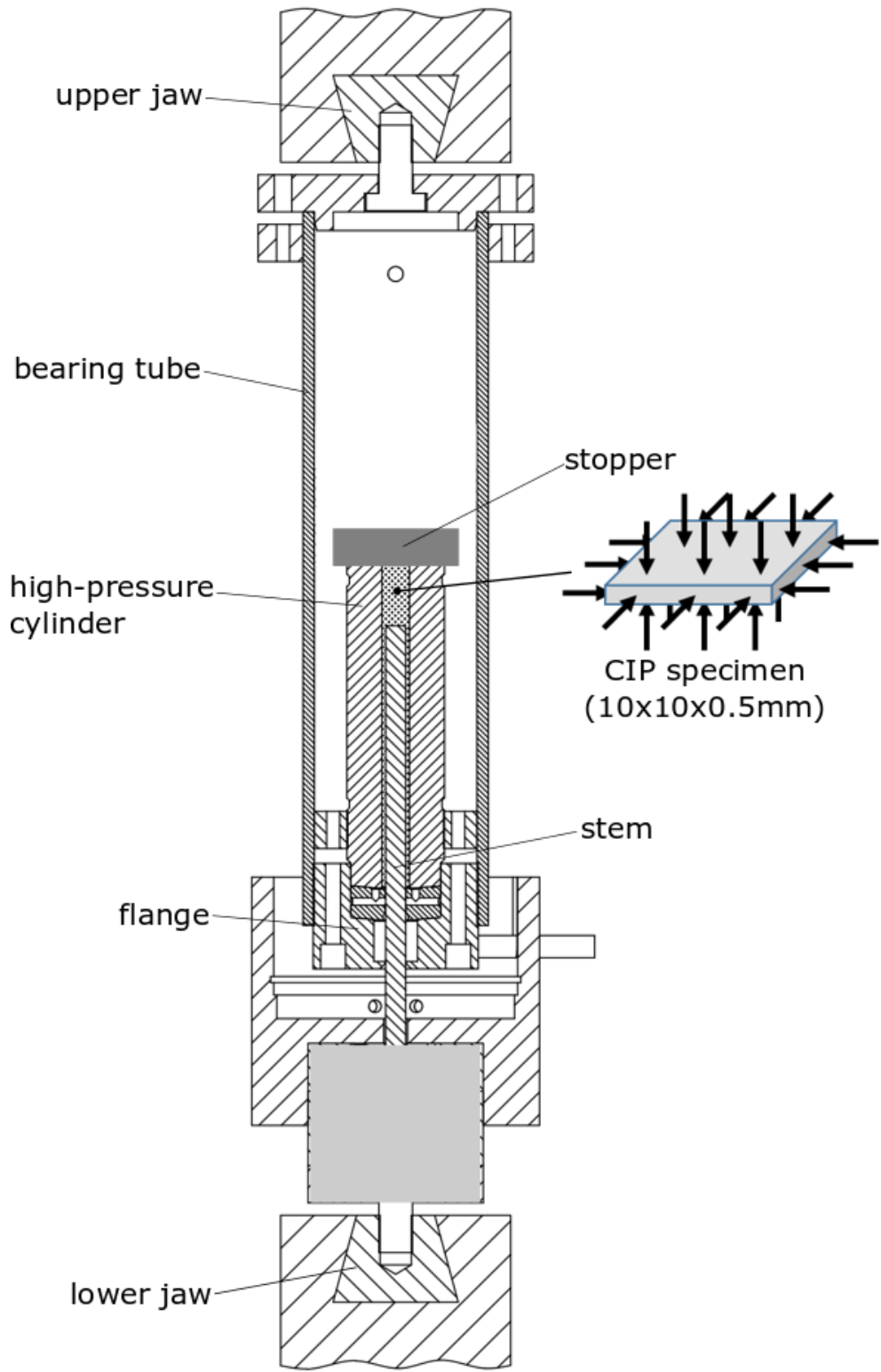

*Figure 3. Schematic description of the device to carry out the high hydrostatic pressure post-processing at room temperature.*

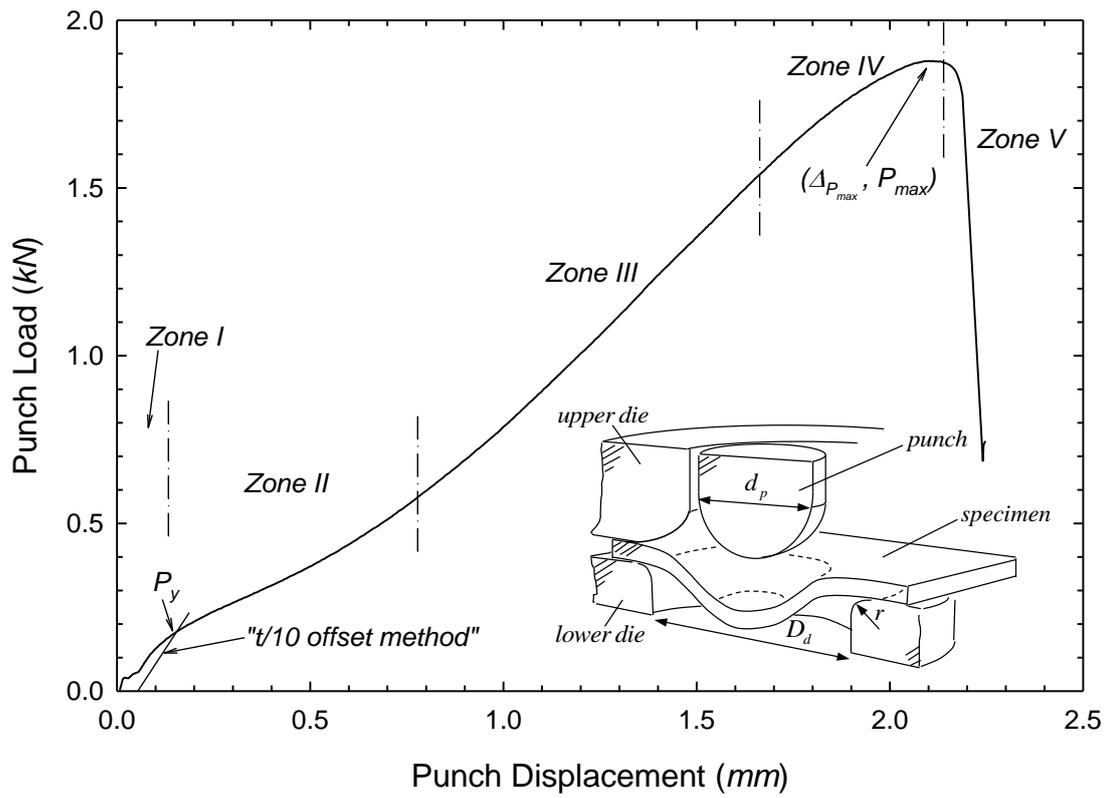

*Figure 4. Typical load-displacement curve and schematic description of the experimental setup involved in the Small Punch Test.*

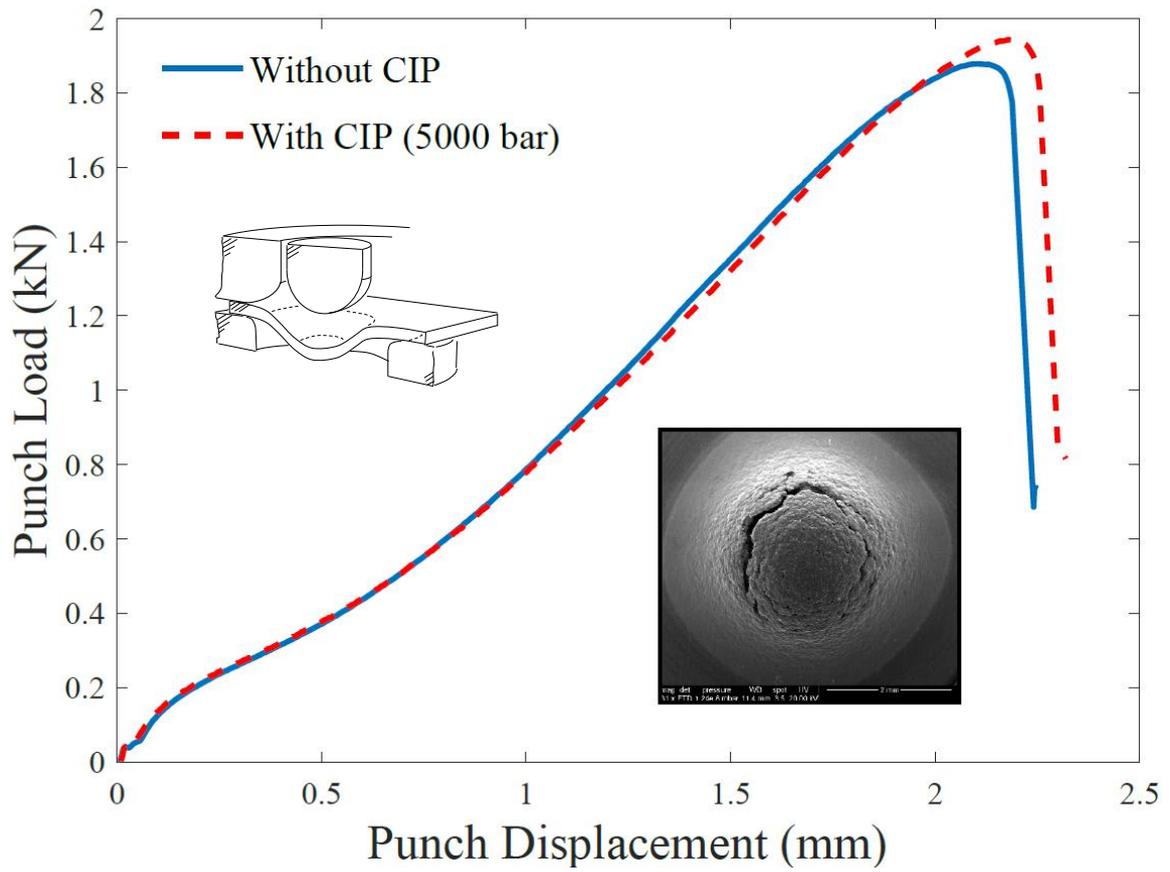

*Figure 5. SPT load-displacement curves (TS orientation) for the AISI 316L with and without CIP.*

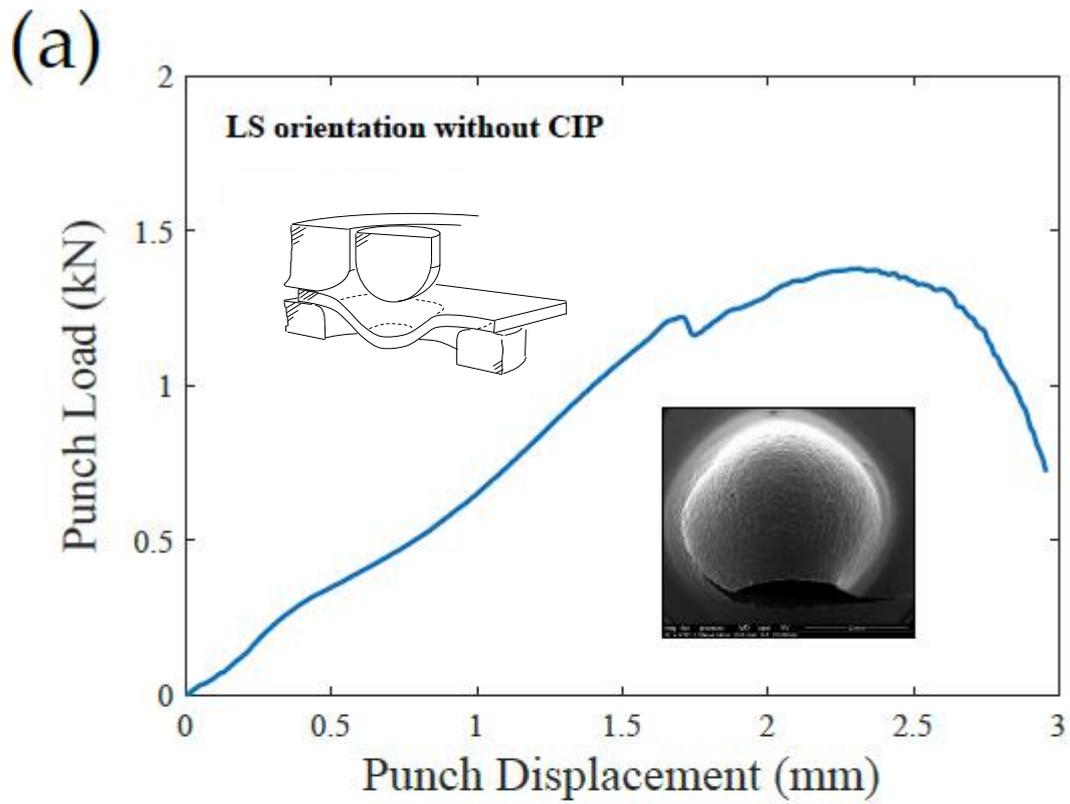

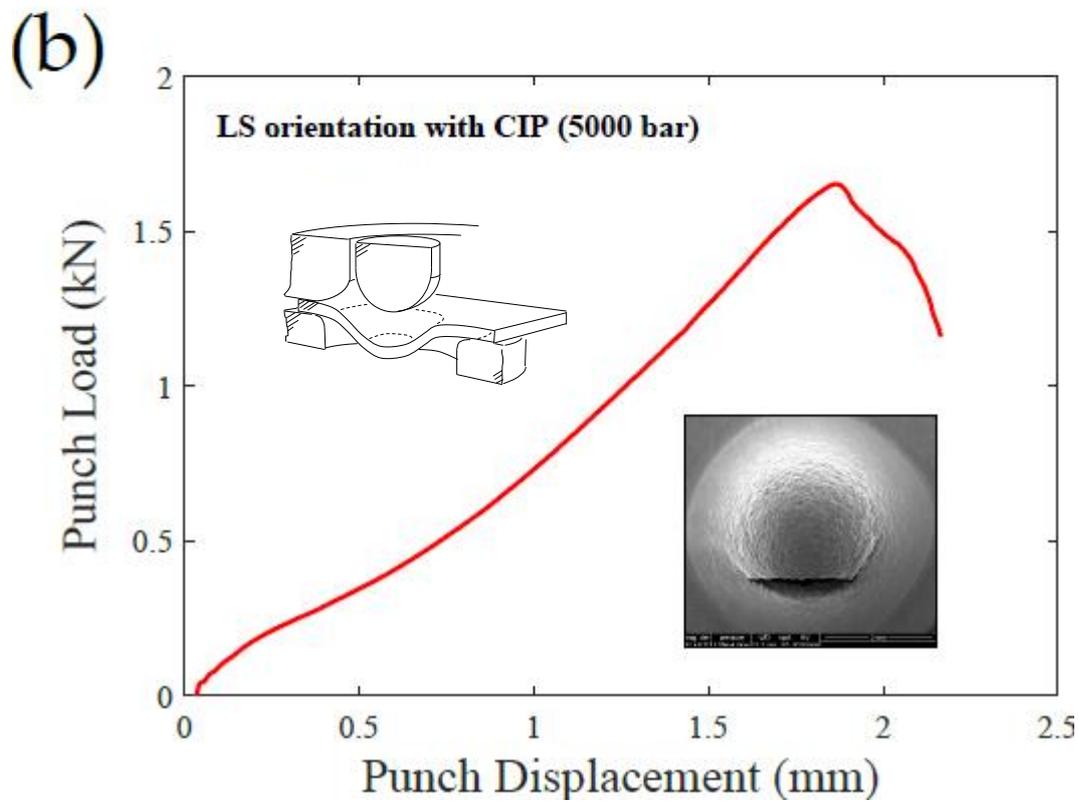

*Figure 6. SPT load-displacement curves for specimens showing premature fracture in L direction: a) B3 specimen and b) E3 specimen.*

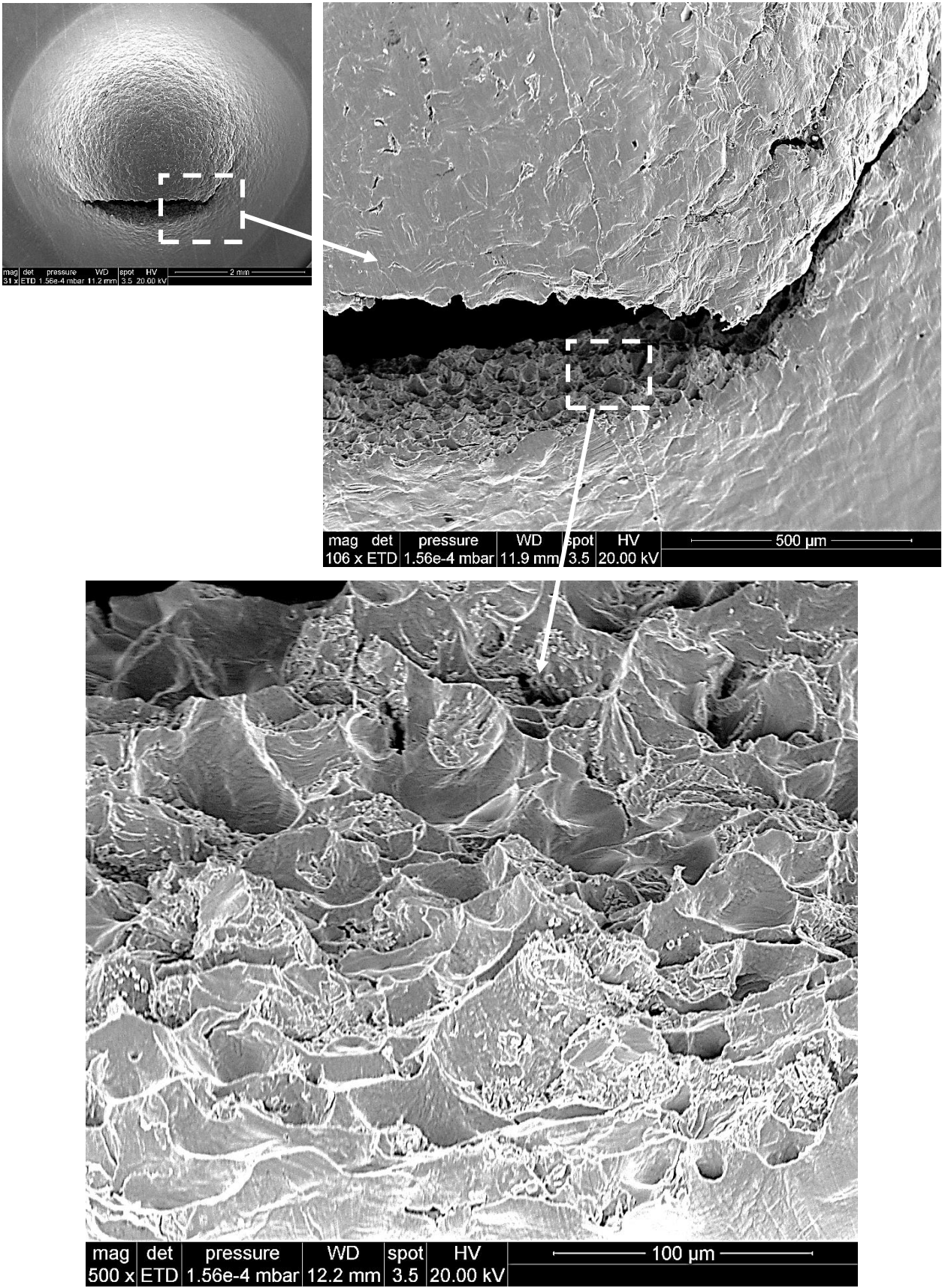

*Figure 7. Aspect of premature fracture (ductile-intergranular) in the SPT specimen E3.*

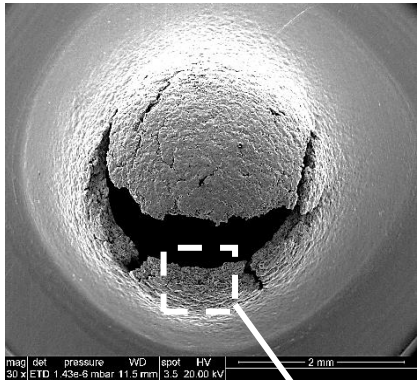
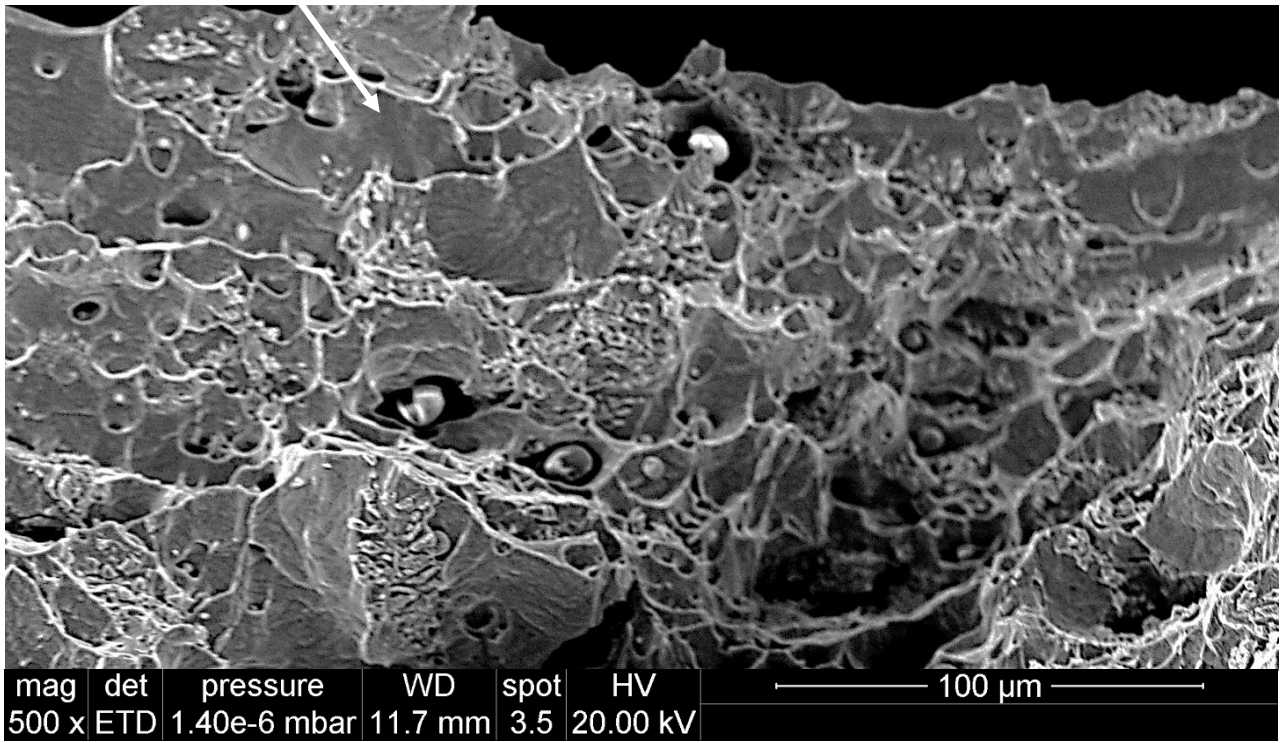

*Figure 8. Aspect of typical ductile fracture in SPT specimens.*